\documentclass[epj]{webofc}
\usepackage[utf8]{inputenc}
\usepackage[varg]{txfonts}   
\usepackage{booktabs}
\usepackage{xcolor}
\definecolor{darkred}{rgb}{0.4,0.0,0.0}
\definecolor{darkgreen}{rgb}{0.0,0.4,0.0}
\definecolor{darkblue}{rgb}{0.0,0.0,0.4}
\usepackage[bookmarks,linktocpage,colorlinks,
    linkcolor = darkred,
    urlcolor  = darkblue,
    citecolor = darkgreen]{hyperref}
\usepackage{subfigure}
\usepackage{amsmath}
\usepackage{braket}
\usepackage{tikz}
\wocname{EPJ Web of Conferences}
\woctitle{Lattice2017}
\DeclareRobustCommand{\abs}[1]{\left\lvert #1 \right\rvert }

\DeclareRobustCommand{\order}[1]{\mathcal{O}\left(#1\right)}
\DeclareRobustCommand{\crossproduct}[0]{\boldsymbol\times} 
\DeclareRobustCommand{\cross}[0]{\crossproduct} 
\newcommand{\be}{\begin{equation}}
\newcommand{\ee}{\end{equation}}
\begin{document}
%
\selectlanguage{english}
\title{Background field Landau mode operators for the nucleon}
\author{%
\firstname{Waseem} \lastname{Kamleh}\inst{1}\fnsep\thanks{Speaker, \email{waseem.kamleh@adelaide.edu.au}}
\firstname{Ryan}  \lastname{Bignell}\inst{1}\and
\firstname{Derek B.} \lastname{Leinweber}\inst{1} \and
\firstname{Matthias}  \lastname{Burkardt}\inst{2}\fnsep
}
\institute{%
Special Research Centre for the Subatomic Structure of Matter, University of Adelaide, Australia \and
Department of Physics, New Mexico State University, Las Cruces, NM 88003-001, U.S.A.
}
\abstract{%
The introduction of a uniform background magnetic field breaks three-dimensional spatial symmetry for a charged particle and introduces Landau mode effects. Standard quark operators are inefficient at isolating the nucleon correlation function at nontrivial field strengths. We introduce novel quark operators constructed from the two-dimensional Laplacian eigenmodes that describe a charged particle on a finite lattice. These eigenmode-projected quark operators provide enhanced precision for calculating nucleon energy shifts in a magnetic field. Preliminary results are obtained for the neutron and proton magnetic polarisabilities using these methods.
}
\maketitle
\section{Introduction}\label{intro}

In the presence of a magnetic field $\vec{B},$ for small values of the field strength it is possible to form an expansion for the energy $E$ of a baryon as
\begin{equation}
  E(B) = M + \vec{\mu}\,\cdot\,\vec{B} + \frac{ \abs{q\,B}}{2\,M} - \frac{1}{2}\,4\,\pi\,\beta\,B^2 + \cdots
  \label{eq:eb}
\end{equation}
The magnetic moment $\vec{\mu}$ appears in the coefficient of the linear
term, and the quadratic term contains the magnetic
polarisability $\beta.$ The term proportional to $\abs{q\,B}$
corresponds to the lowest Landau level energy, and is only present for charged hadrons.

The magnetic moments and polarisabilities of the proton and neutron
are of great interest, and lattice QCD calculations provide an
opportunity to compare with experimental values and gain insight into
the underlying physics~\cite{Kossert:2002jc,Griesshammer:2012we,Wang:2013cfp}. In
particular, experimental measurements of nucleon magnetic
polarisabilities remain challenging with considerable uncertainties.
Improvement has been seen in recent years \cite{Myers:2014ace},
providing scope for lattice QCD to make important predictions
regarding these values.
However, performing a precise lattice calculation of baryon magnetic
polarisabilities presents its own challenges. In principle, the method
is simple. We can impose a uniform background magnetic field on a
lattice gauge field ensemble at a variety of (small) field strengths~\cite{Smit:1986fn,Bernard:1982yu,Burkardt:1996vb,Chang:2015qxa}
and fit the resulting energies as a function of field strength to
extract the magnetic moment and polarisability~\cite{Martinelli:1982cb,Tiburzi:2012ks}.

In practice, due to the overlap of intermediate multi-pion states,
baryon correlation functions inherently suffer from a rapidly decaying
signal-to-noise problem. This makes the extraction of the magnetic
polarisability numerically difficult as it appears at second order in
the energy expansion.
Previous work studying the magnetic properties of the neutron using
the background field method on the lattice with standard nucleon
interpolating fields has demonstrated the difficulty in obtaining a
precise estimate of the magnetic polarisability~\cite{Primer:2013pva,Chang:2015qxa}.

In standard lattice QCD calculations, the use of three-dimensional
gauge covariant Gaussian smearing on the quark fields at the source
and/or sink is highly effective at isolating the nucleon ground
state. However, the imposition of a uniform background field
fundamentally alters the physics that is present. The three
dimensional spatial symmetry is broken by the magnetic field, and the
dynamics of QCD are perturbed by the electromagnetic interactions
experienced by the charged quarks.

In the absence of QCD interactions, under the influence of a uniform
magnetic field each individual quark will have an associated Landau
energy proportional to its charge. If we then turn on QCD interactions
the quarks will hadronise, such that in the confining phase the Landau
energy corresponds to that of the composite particle. In particular,
when forming a neutron, the $ddu$ quarks must combine such that the
overall Landau energy vanishes, as there is zero net hadronic charge.
Hence, we see that in the confining phase the QCD and magnetic
interactions compete with each other. This leads us to the idea of
using quark operators on the lattice that capture both of these
forces.

\begin{figure}[tb] 
  \centering
  \includegraphics[width=0.9\textwidth]{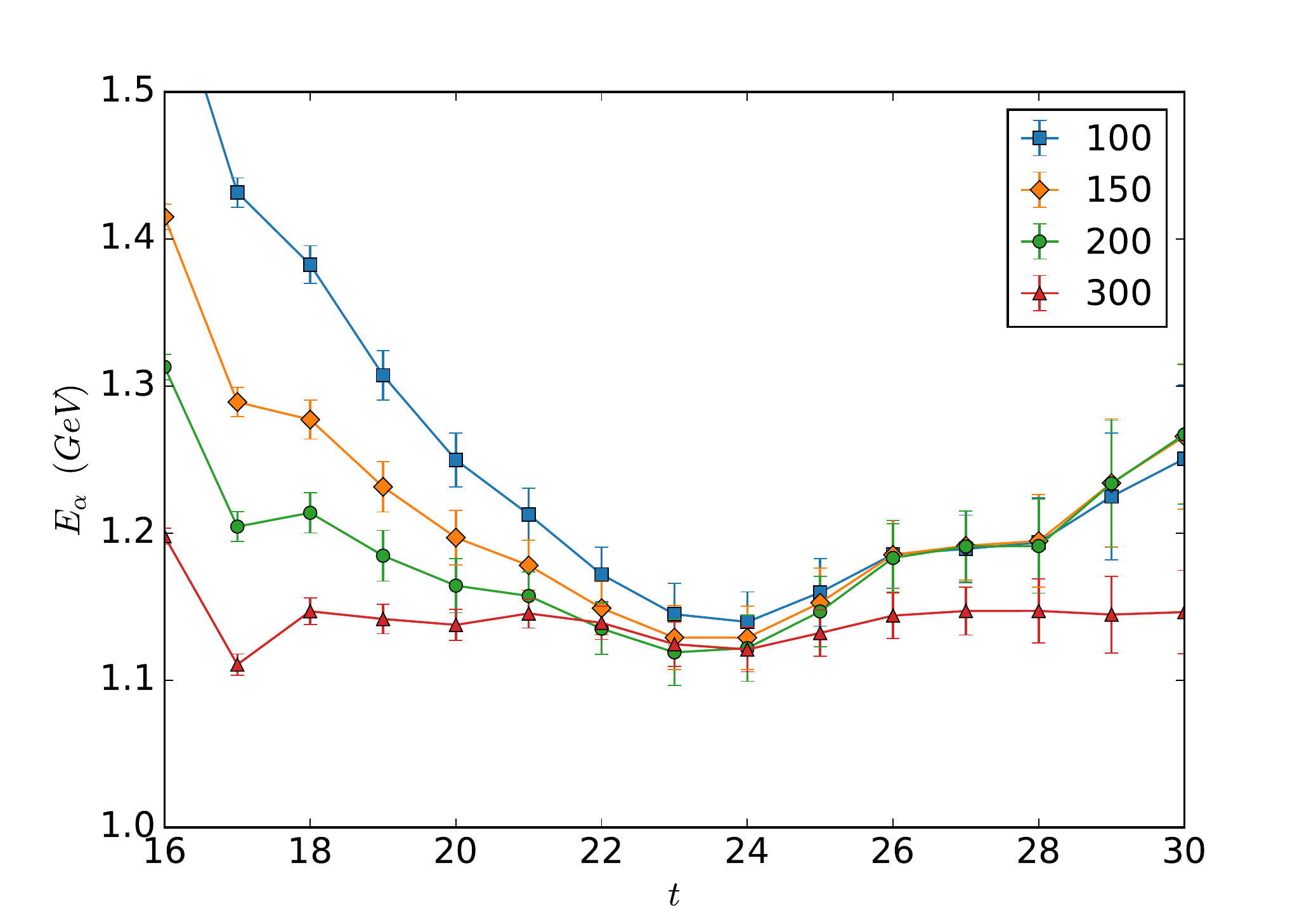}
  \caption{The nucleon effective mass function (at zero magnetic field strength) for various levels of gauge covariant Gaussian smearing at the source, with a point sink.}
  \label{fig:sosmBF0}
\end{figure}

In this work we utilise the freedom of choosing asymmetric source and
sink operators to construct correlation functions that provide better
overlap with the energy eigenstates of the nucleon in a background
magnetic field. At the source, we use 300 sweeps of standard
Gaussian-smearing, tuned to maximise overlap with the nucleon ground
state at $B = 0,$ as shown in Figure~\ref{fig:sosmBF0}. This provides
a representation of the QCD interactions. At the sink, we use a
projection operator based on the eigenmodes of the two-dimensional
lattice Laplacian, in an attempt to capture the physics associated
with the magnetic field.

\section{$U(1)$ Landau mode projection}

To construct a uniform magnetic field (in the $z$ direction) on a lattice with spatial extents $(n_x,n_y,n_z)$, there are multiple gauge-equivalent choices for the potential. We fix the $U(1)$ gauge field to
\begin{align*}
  U_1(x,y) &= \exp(-iB y), \\
  U_2(x,y) &= \left\{ \begin{matrix} 1, & y < n_y-1, \\ \exp(+i B n_y x), & y = n_y-1. \end{matrix} \right.
\end{align*}
Applying the periodic boundary conditions at the corner of the lattice
imposes a quantisation condition,
\be q\,B\,a^2 = \frac{2\,\pi\,k}{n_x\,n_y}, \ee
where the integer $k$ enumerates the magnetic field strength. For a charged scalar particle the lattice Landau levels correspond to the eigenmodes of the 2D Laplacian,
\be \Delta_{\vec{x},\vec{x}\!\:{}'} = 4\delta_{\vec{x},\vec{x}\!\:{}'} - \sum_{\mu=1,2}(U_\mu(\vec{x})\delta_{\vec{x}+\hat{\mu},\vec{x}\!\:{}'} + U^\dag_\mu(\vec{x}-\hat{\mu})\delta_{\vec{x}-\hat{\mu},\vec{x}\!\:{}'}). \label{eq:lap2d} \ee
While in the continuum, each Landau level has an infinite degeneracy, the
degeneracy of the lattice Landau modes is finite and dependent on the
magnetic field-strength. Of primary interest is the lowest Landau
level on the lattice, which has degeneracy equal to the flux quanta
$|k|.$

We first consider the Landau level effects on the lattice as they
apply to hadrons, being colour singlet states. A hadron of (integer)
charge $q$ experiences a field strength $k_B = -3 q \,k_d$ relative to that of the $d$
quark. Consider the momentum-projected hadronic two-point correlator,
\be G(t,\vec{p})=\sum_{\vec{x}}e^{-i\vec{p}\cdot\vec{x}}\,\langle{\Omega}\vert T \{ \chi (t,\vec{x})\,\bar\chi(0)\} \vert{\Omega}\rangle. \ee
Due to the presence of the magnetic field, the energy eigenstates of a
charged hadron (such as the proton) in a magnetic field cannot also be
eigenstates of the $p_x, p_y$ momentum components. Instead of
performing a 3-dimensional Fourier projection, we can instead project
the $x,y$ dependence of $G$ onto the lowest Landau level, and also
select a specific value for the $z$ component of momentum,
\be G(t,\vec{B},p_z)=\sum_{\vec{x}}\psi_{\vec{B}}(x,y)\,e^{-ip_zz}\,\langle{\Omega}\vert T \{ \chi (t,\vec{x})\,\bar\chi(0)\} \vert{\Omega}\rangle. \ee
In the infinite volume case, the lowest Landau mode takes a Gaussian
form, $\psi_{\vec{B}}(x,y) \sim e^{-|qB|\,(x^2+y^2)/4}.$ As has been
noted elsewhere~\cite{Tiburzi:2012ks,Bignell:2017lnd}, in a finite
volume the periodicity alters the form of the wave function. On the
lattice, we can calculate the eigenmodes of the 2D
Laplacian in Eq.~(\ref{eq:lap2d}) and project the correlator for a charged
hadron onto the space spanned by the modes $\psi_{i, \vec{B}}$
associated with the lowest lattice Landau level,
\be G(t,\vec{B},p_z)=\sum_{\vec{x}}\sum_{i=1}^{|3qk_d|}\,\psi_{i,\vec{B}}(x,y)\,e^{-ip_zz}\langle{\Omega}\vert T \{ \chi (t,\vec{x})\,\bar\chi(0)\} \vert{\Omega}\rangle. \label{eq:HadProj}\ee
While this hadronic projection would be applicable to the proton, the
neutron has zero charge $q=0,$ and does not have hadronic Landau
levels. Hence, the energy eigenstates of the neutron can also be
eigenstates of $\vec{p}.$ However, the neutron is composed of quarks,
which due to their charge would have individual Landau levels in the
absence of QCD interactions. This motivates us to try and capture the
Landau mode effects at the quark level.
\begin{figure}[t] 
  \centering
  \includegraphics[page=3,width=0.9\textwidth]{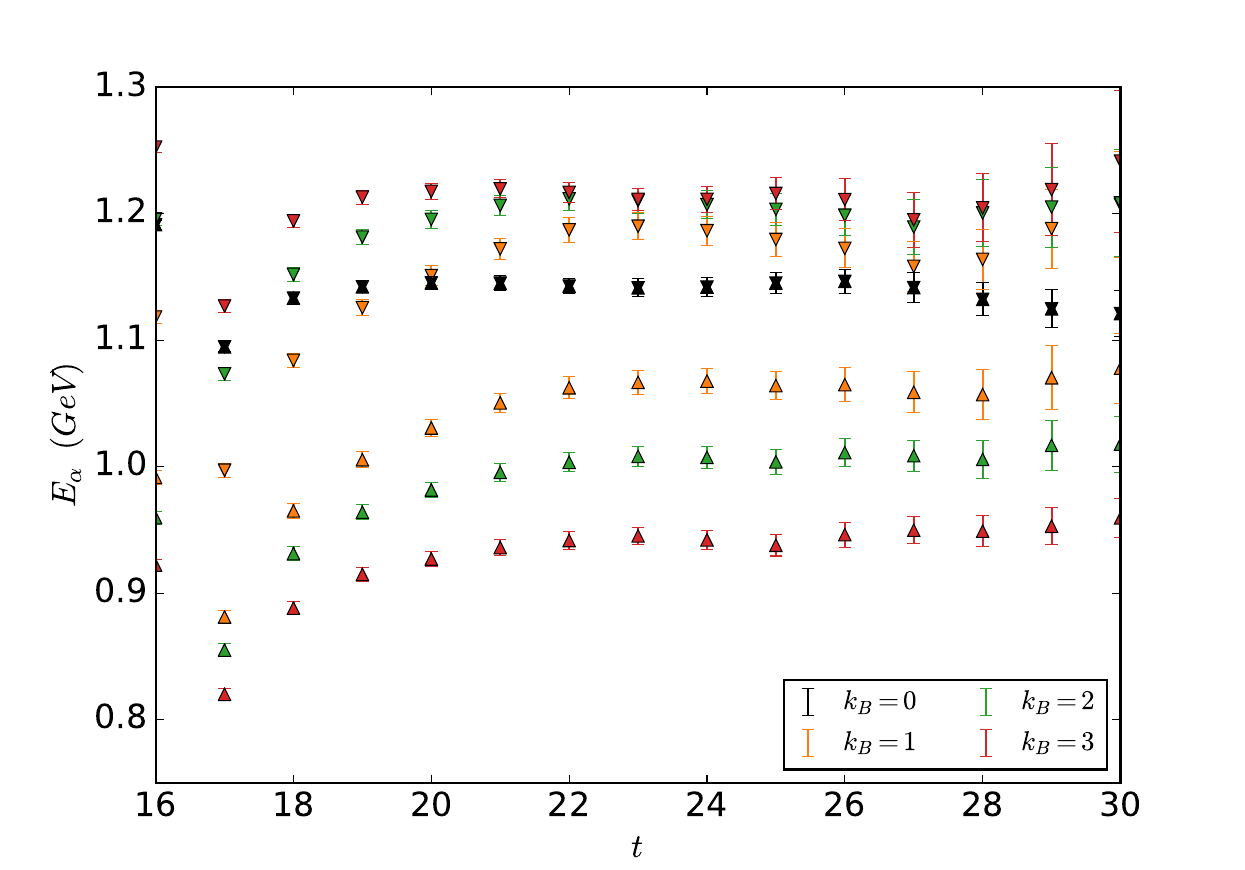}
  \caption{The neutron effective energy function for different magnetic field strengths (enumerated by the integer $k_B).$ Results are shown for both spin polarisations, with the upward/downward pointing triangles corresponding to the spin being aligned/anti-aligned with the magnetic field respectively. We use 300 sweeps of gauge covariant Gaussian smearing at the source. After fixing to Landau gauge, the quark propagators are projected onto the lowest $n$ eigenmodes of the QED-only 2D Laplacian at the sink, where $n$ is equal to $k_B$ for the down quark and $2k_B$ for the up quark. We take advantage of the symmetry in the $z$ direction by averaging over both the positive and negative magnetic fields.}
  \label{fig:neutronEffE}
\end{figure}
We define a projection operator onto lowest $n$ eigenmodes $|\,\psi_{i, \vec{B}}\rangle$ of the 2D Laplacian, 
\be P_n = \sum_{i=1}^{n}\, |\,\psi_{i, \vec{B}}\rangle\,\langle\psi_{i, \vec{B}} \,|\,. \ee
The 2-dimensional projection operator is applied at the sink to the quark propagator.
\be S_n(\vec{x},t;\vec{y},0) = \sum_{\vec{x}\!\:{}'} P_n(\vec{x},\vec{x}\!\;{}')\,S(\vec{x}\!\;{}',t;\vec{y},0)\,. \ee
where $n=\abs{3\,q_f\,k_d}$ modes for the lowest Landau level. As the
$U(1)$ Laplacian is not QCD gauge covariant, for the lattice Landau
mode projection we fix the gluon field to Landau gauge and apply the
appropriate gauge rotation to the quark propagator before
projecting. We are free to do this as the hadronic correlation
function is gauge invariant, so using a gauge fixed sink operator can
only affect the overlap with the ground state, potentially improving
the final precision of our result. Figure~\ref{fig:neutronEffE} shows
the neutron effective energy function using a smeared source and
Landau mode sink for the first 3 non-trivial magnetic field strengths,
along with the smeared source to point sink correlator at $k_B=0$ for
comparison. Our results are calculated using the PACS-CS 2+1 flavour
dynamical lattices~\cite{Aoki:2008sm}, with $m_\pi = 413\text{ MeV}.$ The
calculation is electro-quenched, such that only the valence quarks feel
the magnetic field.

\begin{figure}[t] 
  \centering
  \includegraphics[width=0.9\textwidth]{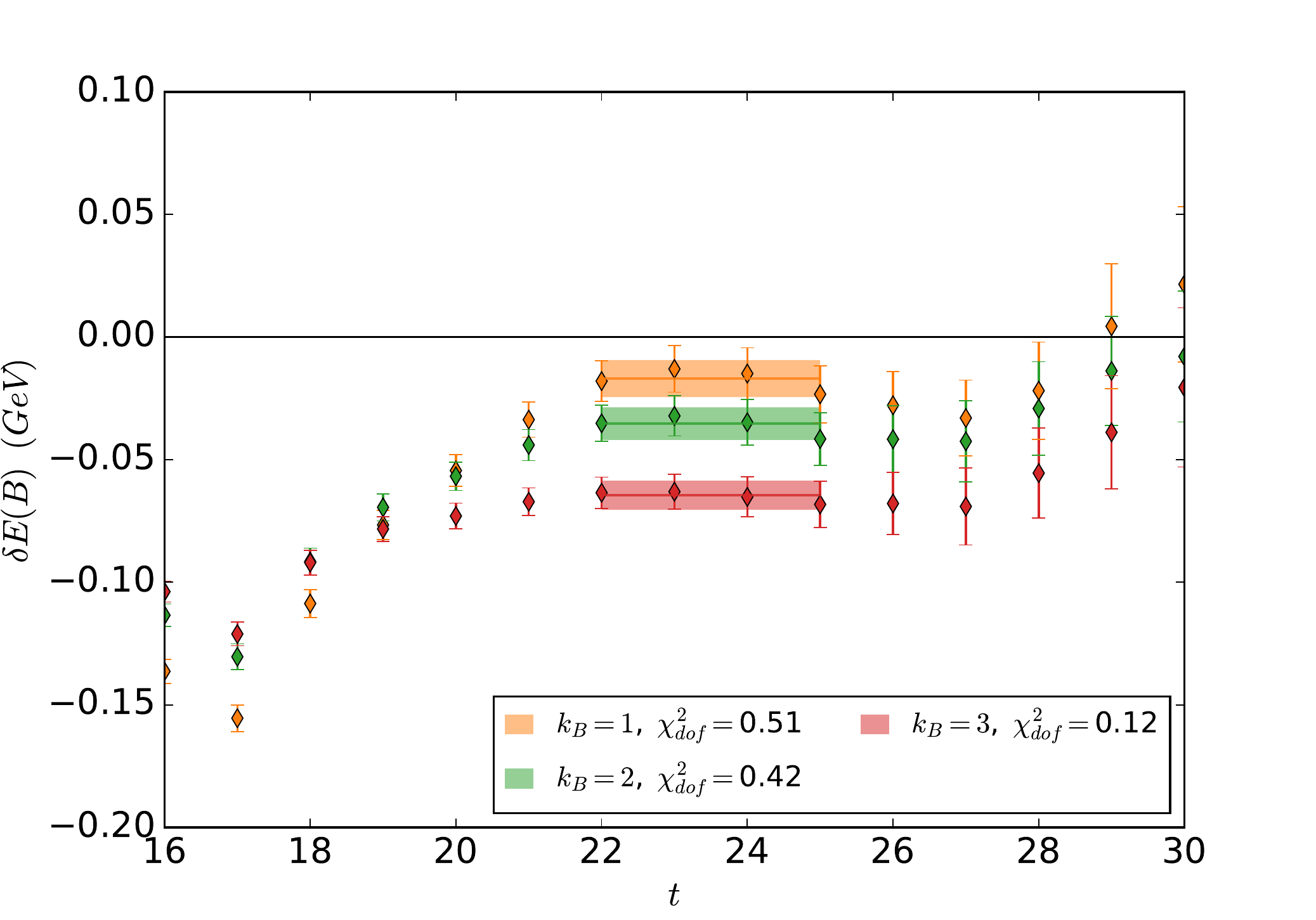}
  \caption{The spin-averaged neutron effective energy shift as a function of Euclidean time (in lattice units), obtained using a smeared source and $U(1)$ Landau mode sink. Results are shown for $k_B = 1,2,3$ (with magnetic field strength increasing away from zero), along with the selected fits and reduced $\chi^2.$ }
  \label{fig:neutronEshiftfits}
\end{figure}

In order to obtain the magnetic polarisability, we take advantage of
the fact that the magnetic moment term cancels when we average over
the two spin polarisations. As the neutron has zero charge and hence
no Landau term in the energy expansion (\ref{eq:eb}), the
polarisibility becomes the leading term in the spin-averaged energy
shift,
\be \delta E(B) = -\frac{1}{2} \, 4\,\pi \,\beta\,B^2 + \order{B^4}. \ee
As the energy shift due to the polarisability is second order in $B,$ a
precise determination of the energy shift is essential. We obtain the
energy shift as a function of field strength $B$ by constructing the
spin-averaged neutron correlator and taking the ratio with the zero
field $B=0$ correlator. Using this ratio it is possible to reduce the
statistical noise in the energy shift, if there are correlated
fluctuations that cancel out. We also take advantage of the symmetry
in the field direction and average over $\pm B$ fields, such that
spin-averaged energy shift correlator $A(B,t)$ at field strength $B$
is given by the following expression
\begin{align}
A(B,t) = \left( \frac{G_{\uparrow}(+B,t) + G_{\downarrow}(-B,t)  }{ G_{\uparrow}(0,t) + G_{\downarrow}(0,t)} \right)
&\times \left( \frac{G_{\downarrow}(+B,t) + G_{\uparrow}(-B,t)  }{ G_{\downarrow}(0,t) + G_{\uparrow}(0,t)} \right).
\label{eq:eratio}
\end{align}
The energy shift is then obtained in the standard way by fitting a constant to the effective energy function
\be \delta E(B,t) = \frac{1}{2} \, \frac{1}{\delta t} \log \frac{A(B,t)}{A(B,t+\delta t)}. \ee
Figure~\ref{fig:neutronEshiftfits} shows our fits for the neutron
spin-averaged energy shift with a smeared source and Landau mode
sink. For the first time, we are able to obtain clear plateaus to fit
this difficult to obtain second-order term. A fit of the resulting
energy shifts as a function of $B$ yields a neutron magnetic polarisability
value of $\beta_n = 1.39(15) \cross 10^{-4}\text{ fm}^3.$

\begin{figure}[t] 
  \centering
  \includegraphics[page=3,width=0.9\textwidth]{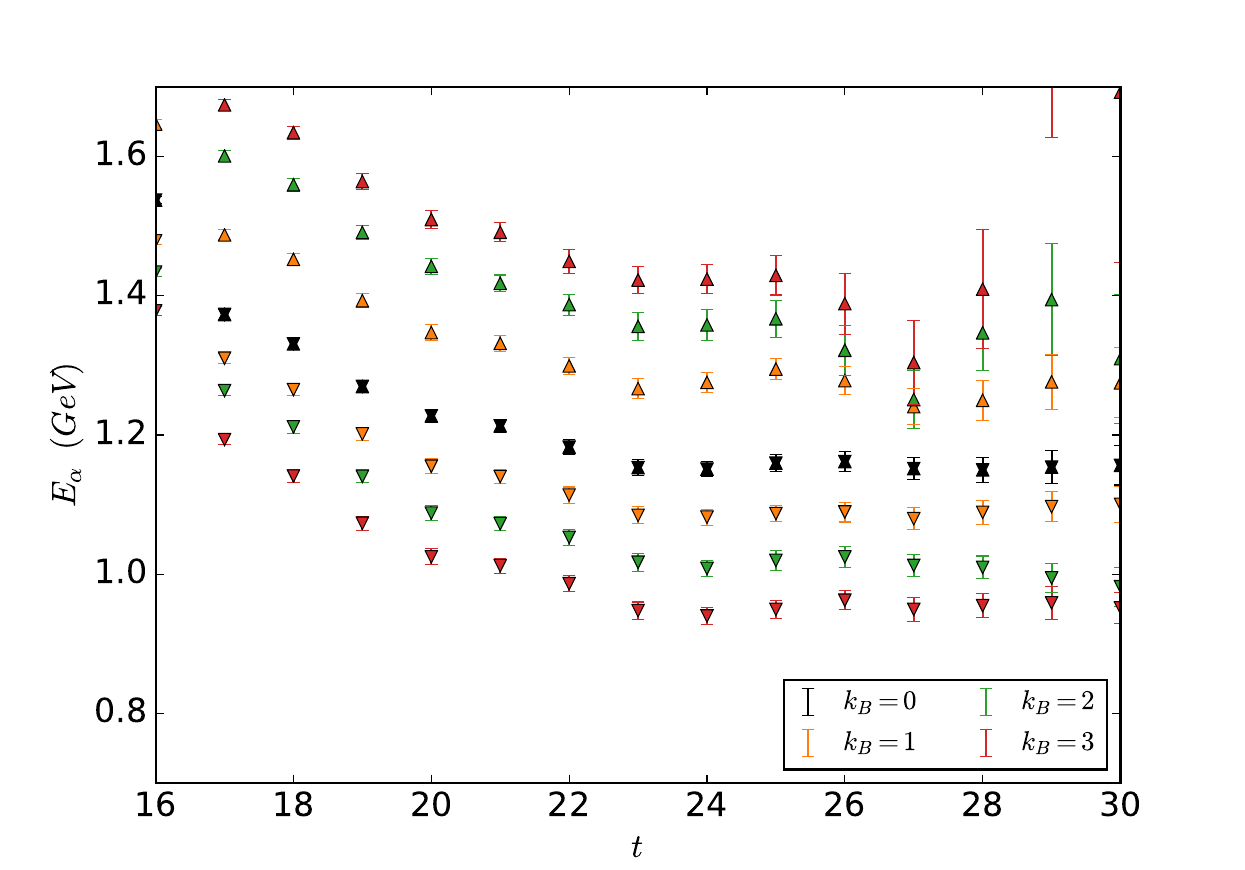}
  \caption{The proton effective energy function for different magnetic field strengths (enumerated by the integer $k_B).$ Results are shown for both spin polarisations, with the upward/downward pointing triangles corresponding to the spin being aligned/anti-aligned with the magnetic field respectively. We use 100 sweeps of gauge covariant Gaussian smearing at the source. The up and down quark propagators are projected onto the lowest $n = 96$ eigenmodes of the QED+QCD 2D Laplacian at the sink. We take advantage of the symmetry in the $z$ direction by averaging over both the positive and negative magnetic fields.}
  \label{fig:protonEffE}
\end{figure}

\section{$U(1)\times S\!U(3)$ Laplacian eigenmode projection}

We can construct a gauge covariant projection operator by
incorporating the $S\!U(3)$ gauge links in the definition of the 2D
Laplacian in Eq.~(\ref{eq:lap2d}). This ensures that the eigenmodes
are QCD covariant, so there is no need to gauge fix. The projector
$P_n$ is then simply a truncation of the completeness relation,
\be 1 = \sum_i |\psi_i\rangle \langle \psi_i | \ee
This truncation has a similar effect to performing (2D) smearing, by
filtering out the high frequency modes. Indeed, we find that for small
values of $n$ the projected hadron correlator becomes noisy, just as
it does when performing large amounts of sink smearing.

The introduction of the QCD interactions into the Laplacian causes the
$U(1)$ modes associated with the different Landau levels to mix, such
that it is no longer possible to clearly identify the modes associated
with the lowest Landau level at small field strengths.  Instead, we
simply choose a fixed number $n > |qk|$ modes to project, where $n$
should be sufficiently large as to avoid introducing large amounts of
noise into the two-point correlation function.

\begin{figure}[t] 
  \centering
  \includegraphics[width=0.9\textwidth]{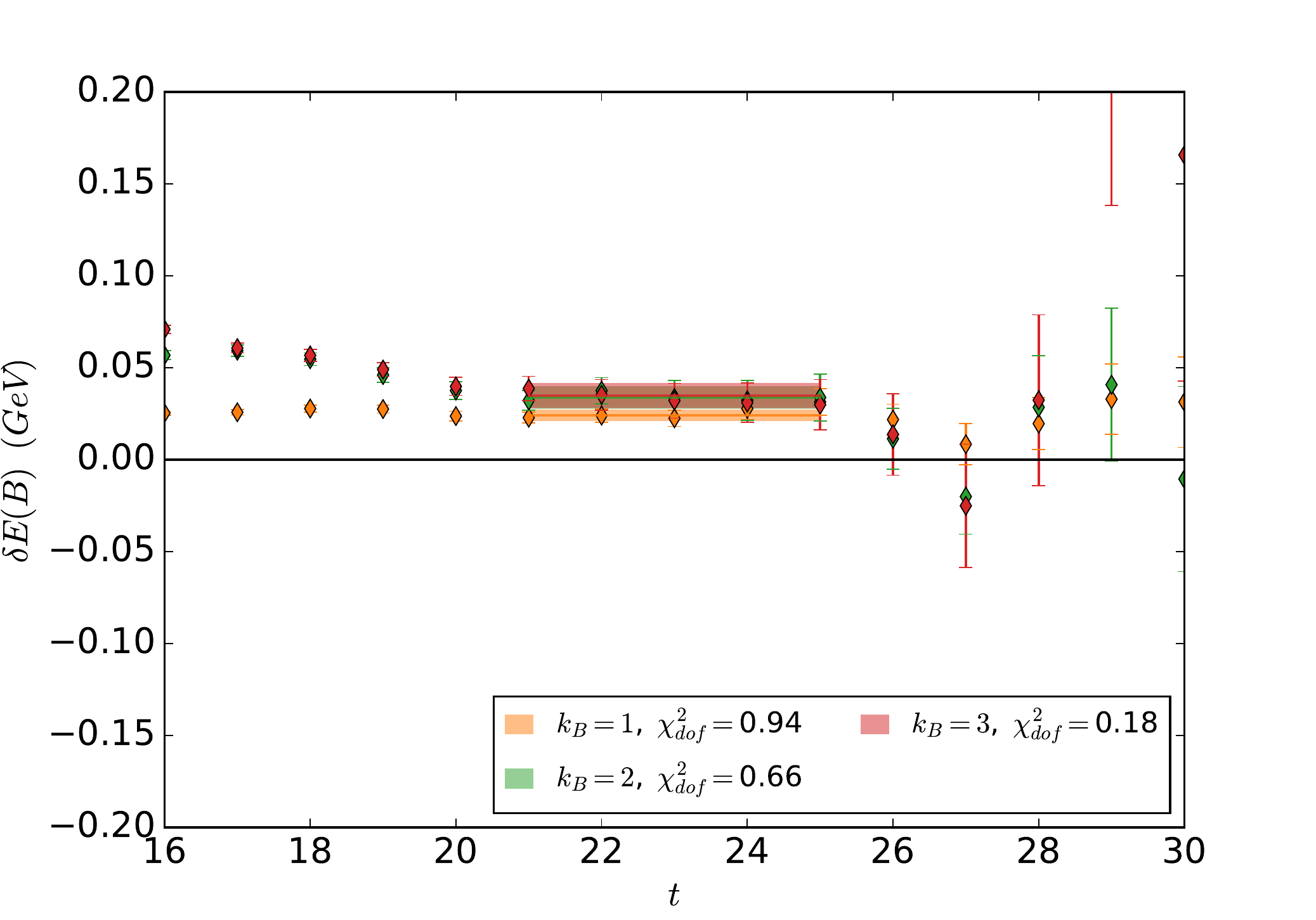}
  \caption{The spin-averaged proton effective energy shift as a function of Euclidean time (in lattice units), obtained using a smeared source and $U(1)\times S\!U(3)$ eigenmode sink. Results are shown for $k_B = 1,2,3$ (with magnetic field strength increasing away from zero), along with the selected fits and reduced $\chi^2.$} 
  \label{fig:protonEshiftfits}
\end{figure}

An advantage of using the $U(1)\times S\!U(3)$ Laplacian projector is
that it is well defined at zero magnetic field strength, where the
$U(1)$ field is equal to unity. This means that the fluctuations at
finite $B$ and $B=0$ are strongly correlated, such that they cancel
out when taking the ratio of the correlators in Eq.~(\ref{eq:eratio}),
providing an improved signal in comparison to the $U(1)$ projection.
This improvement does come at a computational cost, as the
$U(1)\times S\!U(3)$ Laplacian eigenmodes must be calculated on every
configuration, whereas the pure $U(1)$ Laplacian eigenmodes need only
be calculated once.

As the proton is charged, in addition to the quark level $U(1)\times
S\!U(3)$ Laplacian projection applied to the quark propagator, we also
project the proton correlation function onto the lowest hadronic Landau level
using Eq.~(\ref{eq:HadProj}). Furthermore, the spin-averaged
proton energy shift must now include the hadronic Landau energy term
along with the second order polarisability term,
\be \delta E(B) = \frac{ \abs{q\,B}}{2\,M} - \frac{1}{2}\,4\,\pi \,\beta\,B^2 + \order{B^4}. \ee
Figure~\ref{fig:protonEshiftfits} shows our fits for the proton
spin-averaged energy shift with a smeared source and $U(1)\times
S\!U(3)$ Laplacian eigenmode sink. Due to the Landau and polarisibility
terms having opposite sign, the variation in the energy shift as a
function of field strength is much smaller than for the
neutron. Nonetheless, we are again able to obtain clear plateaus,
enabling us to extract the energy shifts and fit them as a function of
field strength $B,$ providing a preliminary result for the proton
magnetic polarisability of $\beta_p = 1.15(24) \cross 10^{-4}\text{ fm}^3.$

\section{Summary}

We have studied the energy shifts of the neutron and proton induced by
a magnetic field using a smeared source and two different Laplacian
eigenmode projectors at the sink. For the neutron, gauge fixing the
quark propagators and projecting on the lowest Landau level using
$U(1)$ eigenmodes provides us with clear plateaus for the
spin-averaging energy shifts. For the proton, similar success is
obtained by a gauge covariant projector formed from the eigenmodes of
the $U(1)\times S\!U(3)$ Laplacian.

It is worthwhile to consider for a moment the physics that underpins
the success of our Laplacian eigenmode projection methods. As we are
interested in obtaining the magnetic polarisability, we want to work
with small $\vec{B}$ field strengths for the perturbative energy expansion
to be valid; hence we are in the confined phase of QCD. Confined
quarks cannot have individual Landau levels. Nonetheless, the effects
of the magnetic field on the quark distribution in the nucleon appear
to be significant. A possible explanation for this may be linked to
the chiral magnetic effect, induced by the interplay between the
magnetic field and the local gauge field topology. Indeed, based on
topological arguments in two-dimensions, a recent staggered quark
study at finite temperature finds that the contribution of the
lowest Landau level eigenmodes remains important even after QCD
interactions are introduced~\cite{Bruckmann:2017pft}.

Based on the success of the methods presented herein, we are now in a
position to provide precise determinations of the nucleon magnetic
polarisabilities using lattice QCD, and our calculations will be
published in full elsewhere~\cite{in-prep}.

%


\end{document}